\documentstyle[sprocl,psfig,epsf]{article}

\def\be{\begin{equation}}
\def\ee{\end{equation}}
\def\bea{\begin{eqnarray}}
\def\eea{\end{eqnarray}}

\begin{document}

\begin{flushright}
\begin{tabular}{l}
UCLA/00/TEP/03
\end{tabular}
\end{flushright}
 
\vspace{8mm}

\begin{center}
 
{\Large \bf Cosmology of SUSY Q-balls}\footnote{Invited talk 
  at the International Workshop on Particle Physics and the Early Universe
  ({\bf COSMO-99}), ICTP, Trieste, Italy, September 27 - October 2, 1999.}

\vspace{8mm}

{\large
Alexander Kusenko}\footnote{ email address: kusenko@physics.ucla.edu}
 
\vspace{6mm}
Department of Physics and Astronomy, University of California, Los
Angeles, CA 90095-1547  \\
and  \\
RIKEN BNL Research Center, Brookhaven National
Laboratory, Upton, NY 11973, USA

\vspace{12mm} 
 
{\bf Abstract}%

\end{center}

Supersymmetric extensions of the Standard Model
  predict the existence of Q-balls, some of which can be entirely stable.
  Both stable and unstable Q-balls can play an important role in cosmology.
  In particular, Affleck--Dine baryogenesis can result in a copious
  production of stable baryonic Q-balls, which can presently exist as a
  form of dark matter.  Formation and decay of unstable Q-balls can also
  have some important effects on baryogenesis and phase transitions.
 
\vfill
 
\pagestyle{empty}
 
\pagebreak
 
\pagestyle{plain}
\pagenumbering{arabic}

\title{Cosmology of SUSY Q-balls
}

\author{Alexander Kusenko}

\address{Department of Physics and Astronomy, University of California, Los
Angeles, CA 90095-1547 
\\E-mail: kusenko@ucla.edu} 


\maketitle\abstracts{Supersymmetric extensions of the Standard Model
  predict the existence of Q-balls, some of which can be entirely stable.
  Both stable and unstable Q-balls can play an important role in cosmology.
  In particular, Affleck--Dine baryogenesis can result in a copious
  production of stable baryonic Q-balls, which can presently exist as a
  form of dark matter.  Formation and decay of unstable Q-balls can also
  have some important effects on baryogenesis and phase transitions. }

\section{Non-topological solitons in MSSM} 

In a class of theories with interacting scalar fields $\phi$ that carry
some conserved global charge, the ground state is a Q-ball~\cite{q}, a lump
of coherent scalar condensate that can be described semiclassically as a 
non-topological soliton of the form  
\begin{equation}
\phi(x,t) = e^{i \omega t} \bar{\phi}(x).
\label{q}
\end{equation}
Q-balls exist whenever the scalar potential satisfies certain conditions
that were first derived for a single scalar degree of freedom~\cite{q} with
some abelian global charge and were later generalized to a theory of many
scalar fields with different charges~\cite{ak_mssm}.  Non-abelian global
symmetries~\cite{nonabelian} and abelian local symmetries~\cite{gauge} can
also yield Q-balls.

It turns out that all phenomenologically viable supersymmetric extensions
of the Standard Model predict the existence of non-topological
solitons~\cite{ak_mssm} associated with the conservation of baryon and
lepton number. If the physics beyond the standard model reveals some
additional global symmetries, this will further enrich the spectrum of
Q-balls~\cite{Demir}.  The MSSM admits a large number of different Q-balls,
characterized by (i) the quantum numbers of the fields that form a
spatially-inhomogeneous ground state and (ii) the net global charge of this
state.

First, there is a class of Q-balls associated with the tri-linear
interactions that are inevitably present in the MSSM~\cite{ak_mssm}.  The
masses of such Q-balls grow linearly with their global charge, which can be
an arbitrary integer number~\cite{ak_qb}.  Baryonic and leptonic Q-balls of
this variety are, in general, unstable with respect to their decay into
fermions.  However, they could form in the early universe through the
accretion of global charge~\cite{gk,ak_pt} or, possibly, in a first-order
phase transition~\cite{s_gen}.

The second class~\cite{dks} of solitons comprises the Q-balls whose VEVs
are aligned with some flat directions of the MSSM.  The scalar field inside
such a Q-ball is a gauge-singlet~\cite{kst} combination of squarks and
sleptons with a non-zero baryon or lepton number.  The potential along a
flat direction is lifted by some soft supersymmetry-breaking terms that
originate in a ``hidden sector'' of the theory at some scale $\Lambda_{_S}$
and are communicated to the observable sector by some interaction with a
coupling $g$, so that $g \Lambda \sim 100$~GeV.  Depending on the strength
of the mediating interaction, the scale $\Lambda_{_S}$ can be as low as a
few TeV (as in the case of gauge-mediate SUSY breaking), or it can be some
intermediate scale if the mediating interaction is weaker (for instance,
$g\sim \Lambda_{_S}/m_{_{Planck}}$ and $\Lambda_{_S}\sim 10^{10}$~GeV in
the case of gravity-mediated SUSY breaking).  For the lack of a definitive
scenario, one can regard $\Lambda_{_S}$ as a free parameter.  Below
$\Lambda_{_S}$ the mass terms are generated for all the scalar degrees of
freedom, including those that parameterize the flat direction.  At the
energy scales larger than $\Lambda_{_S}$, the mass terms turn off and the
potential is ``flat'' up to some logarithmic corrections.  If the Q-ball
VEV extends beyond $\Lambda_{_S}$, the mass of a soliton~\cite{dks,ks} is
no longer proportional to its global charge $Q$, but rather to $Q^{3/4}$.

This allows for the existence of some entirely stable Q-balls with a large
baryon number $B$ (B-balls).  Indeed, if the mass of a B-ball is $M_{_B} \sim
({\rm 1~TeV}) \times B^{3/4}$, then the energy per baryon number
$(M_{_B}/B)\sim ({\rm 1~TeV}) \times B^{-1/4}$ is less than 1~GeV for $B >
10^{12}$.  Such large B-balls cannot  dissociate into protons and neutrons
and are entirely stable thanks to the conservation of energy and the baryon
number.  If they were  produced in the early universe, they would exist at
present as a form of dark matter~\cite{ks}.  

\section{Fragmentation of Affleck--Dine condensate into Q-balls}

Several mechanisms could lead to formation of B-balls and L-balls in the
early universe. First, they can be produced in the course of a phase
transition~\cite{s_gen}.  Second, thermal fluctuations of a baryonic and
leptonic charge can, under some conditions, form a Q-ball.  Finally, a
process of a gradual charge accretion, similar to nucleosynthesis, can take
place~\cite{gk,ak_pt,dew}.  However, it seems that the only process that can
lead to a copious production of very large, and, hence, stable, B-balls, is
fragmentation of the Affleck-Dine condensate~\cite{ks}. 

At the end of inflation, the scalar fields of the MSSM develop some large
expectation values along the flat directions, some of which have a non-zero
baryon number~\cite{ad}. Initially, the scalar condensate has the form
given in eq.~(\ref{q}) with $\bar{\phi}(x)= const$ over the length scales
greater than a horizon size. One can think of it as a universe filled with
Q-matter.  The relaxation of this condensate to the potential minimum is
the basis of the Affleck--Dine (AD) scenario for baryogenesis.

It was often assumed that the condensate remains spatially homogeneous from
the time of formation until its decay into the matter baryons.  This
assumption is, in general, incorrect.  In fact, the initially homogeneous
condensate can become unstable~\cite{ks} and break up into Q-balls whose
size is determined by the potential and the rate of expansion of the
Universe.  B-balls with $12 < \log_{10} B < 30$ can form naturally
from the breakdown of the AD condensate.  These are entirely
stable if the flat direction is ``sufficiently flat'', that is if the
potential grows slower than $\phi^2$ on the scales or the order of
$\bar{\phi}(0)$.   The evolution of the primordial condensate can be
summarized as follows: 

\vspace{3mm}
\psfig{figure=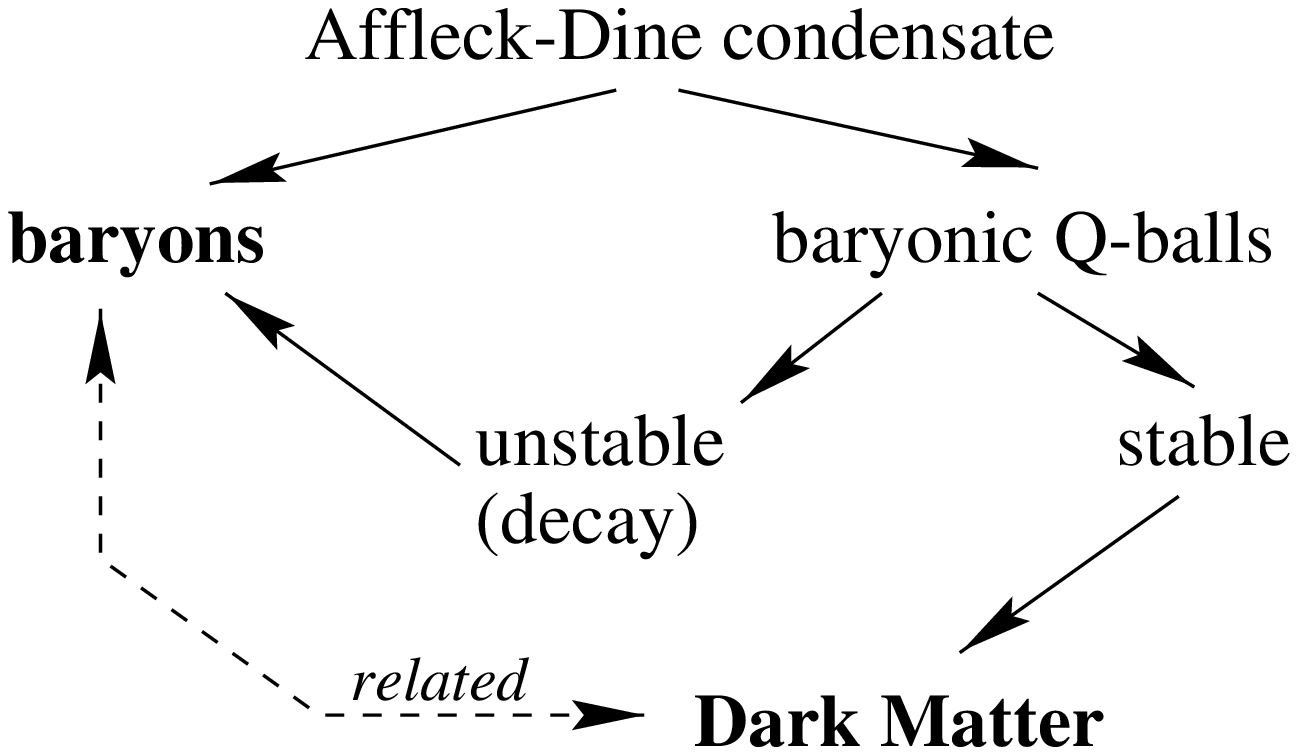,height=1.5in,width=3.5in}

This process has been analyzed analytically~\cite{ks,em} in the linear
approximation.  Recently, some impressive numerical simulations of Q-ball
formation have been performed~\cite{kasuya}; they confirm that the
fragmentation of the condensate into Q-balls occurs in some Affleck-Dine
models.  The global charges of Q-balls that form this way are model
dependent.  The subsequent collisions~\cite{ks,Axenides:1999hs} can further
modify the distribution of soliton sizes.

\section{SUSY Q-balls as dark matter} 

Conceivably, the cold dark matter in the Universe can be made up entirely
of SUSY Q-balls.  Since the baryonic matter and the dark matter share the
same origin in this scenario, their contributions to the mass density of
the Universe are related.  Therefore, it is easy to understand why the
observations find $\Omega_{_{DARK}} \sim \Omega_{B} $ within an order of
magnitude.  This fact is extremely difficult to explain in models that
invoke a dark-matter candidate whose present-day abundance is determined by
the process of freeze-out, independent of baryogenesis.  If this is the
case, one could expect $\Omega_{_{DARK}}$ and $\Omega_{B} $ to be different
by many orders of magnitude.  If one doesn't want to accept this equality
as fortuitous, one is forced to hypothesize some {\it ad hoc}
symmetries~\cite{kaplan} that could relate the two quantities.  In the MSSM
with AD baryogenesis, the amounts of dark-matter Q-balls and the ordinary
matter baryons are related~\cite{ks}.  One predicts~\cite{lsh}
$\Omega_{_{DARK}} = \Omega_{B} $ for B-balls with $B \sim 10^{26}$.  This
size is in the middle of the range of Q-ball sizes that can form in the
Affleck--Dine scenario~\cite{ks,em,kasuya}. 

The value $B\sim 10^{26}$ is well above the present experimental lower
limit on the baryon number of an average relic B-ball, under the assumption
that all or most of cold dark matter is made up of Q-balls.  On their
passage through matter, the electrically neutral baryonic SUSY Q-balls can
cause a proton decay, while the electrically charged B-balls produce
massive ionization.  Although the condensate inside a Q-ball is
electrically neutral~\cite{kst}, it may pick up some electric charge
through its interaction with matter~\cite{kkst}.  Regardless of its ability
to retain electric charge, the Q-ball would produce a straight track in a
detector and would release the energy of, roughly, 10 GeV/mm.  The present
limits~\cite{kkst,exp} constrain the baryon number of a relic dark-matter
B-ball to be greater than $10^{22}$.  Future experiments are expected to
improve this limit.  It would take a detector with the area of several
square kilometers to cover the entire interesting range $B\sim 10^{22}
... 10^{30}$.

The relic Q-balls can accumulate in neutron stars and can lead to their
ultimate destruction over a time period from one billion years to longer
than the age of the Universe~\cite{sw}.  If the lifetime of a neutron star
is in a few Gyr range, the predicted mini-supernova explosions may be
observable.

\section{B-ball baryogenesis}

An interesting scenario that relates the amounts of baryonic and dark matter
in the Universe, and in which the dark-matter particles are produced from
the decay of unstable B-balls was proposed by Enqvist and
McDonald~\cite{em}.  

\section{Phase transitions precipitated by solitosynthesis} 

In the false vacuum, a rapid growth of non-topological
solitons~\cite{gk} can precipitate an otherwise impossible or slow phase
transition~\cite{ak_pt}.  

\begin{figure}
\setlength{\epsfxsize}{2.5in}
\setlength{\epsfysize}{1.8in}
\centerline{\epsfbox{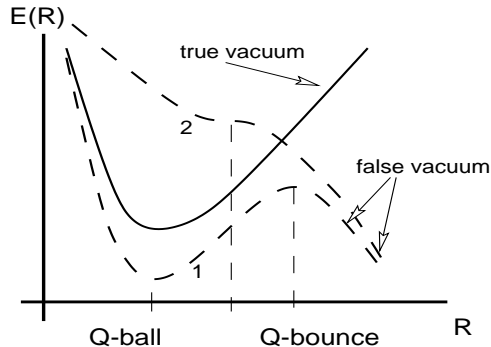}}
\caption{
Energy (mass) of a soliton as a function of its size.  In the true vacuum, 
Q-ball is the global minimum of energy (solid curve).  In the false vacuum,
if the charge is less than some critical value, there are two solutions: a
``stable'' Q-ball, and an unstable ``Q-bounce'' (dashed curve 1) .  In the
case of a critical charge (curve 2), there is only one solution, which is
unstable. 
} 
\label{fig2}
\end{figure}

Let us suppose the system is in a metastable false vacuum that preserves
some U(1) symmetry.  The potential energy in the Q-ball interior is
positive in the case of a true vacuum, but negative if the system is in the
metastable false vacuum. In either case, it grows as the third power of the
Q-ball radius $R$.  The positive contribution of the time derivative to the
soliton mass can be written as $Q^2/\int \bar{\phi}^2(x)d^3x \propto
R^{-3}$, and the gradient surface energy scales as $R^2$.  In the true
vacuum, all three contributions are positive and the Q-ball is the absolute
minimum of energy (Fig.~\ref{fig2}).  However, in the false vacuum, the
potential energy inside the Q-ball is negative and goes as $\propto -R^3$.
As shown in Fig.~\ref{fig2}, for small charge $Q$, there are two stationary
points, the minimum and the maximum.  The former corresponds to a Q-ball
(which is, roughly, as stable as the false vacuum is), while the latter is
a critical bubble of the true vacuum with a non-zero charge.

There is a critical value of charge $Q=Q_c$, for which the only stationary
point is unstable.  If formed, such an unstable bubble will expand.

If the Q-ball charge increases gradually, it eventually reaches the
 critical value.  At that point Q-ball expands and converts space into a
 true-vacuum phase.  In the case of tunneling, the critical bubble is
 formed through coincidental coalescence of random quanta into an extended
 coherent object.  This is a small-probability event.  If, however, a
 Q-ball grows through charge accretion, it reaches the critical size with
 probability one, as long as the conditions for growth~\cite{ak_pt} are
 satisfied.  The phase transition can proceed at a much faster rate than it
 would by tunneling.

\section{Conclusion}

Supersymmetric models of physics beyond the weak scale offer two plausible 
candidates for cold dark matter.  One is the lightest supersymmetric
particle, which is stable because of R-parity.  Another one is a
stable non-topological soliton, or Q-ball, carrying some baryonic charge. 

SUSY Q-balls make an appealing dark-matter candidate because their
formation is a natural outcome of Affleck--Dine baryogenesis and requires
no unusual assumptions. 

In addition, formation and decay of unstable Q-balls can have a dramatic
effect on baryogenesis, dark matter, and the cosmic microwave
background.  Production of unstable Q-balls in the false vacuum can cause an
unusually fast first-order phase transition.

\end{document}